# Unveiling the electronic structure of pseudo-tetragonal WO$_3$ thin films

F. Mazzola[1,2,*], H. Hassani[3,4,*], D. Amoroso[3], S.K. Chaluvadi[2], J. Fujii[2], V. Polewczyk[2], P. Rajak[2], Max Koegler[2], R. Ciancio[5] B. Partoens[4], G. Rossi[2,6], 1 I. Vobornik[2], P. Ghosez[3], and P. Orgiani[2]

1 Department of Molecular Sciences and Nanosystems, Ca' Foscari University of Venice, 30172 Venice, Italy
2 Istituto Officina dei Materiali (IOM)-CNR, Area Science Park, S.S.14, Km 163.5, 34149 Trieste, Italy
3 Theoretical Materials Physics, Q-MAT, CESAM, Université de Liège, B-4000 Liège, Belgium
4 Department of Physics, University of Antwerp, Groenenborgerlaan 171, 2020 Antwerp, Belgium
5 Area Science Park, Padriciano 99, 34149 Trieste, Italy
6 University of Milano, I-20133 Milano, Italy

(Dated: April 11, 2023)

WO$_3$ is a binary *5d* compound which has attracted remarkable attention due to the vast array of structural transitions that it undergoes in its bulk form. In the bulk, a wide range of electronic properties has been demonstrated, including metal-insulator transitions and superconductivity upon doping. In this context, the synthesis of WO$_3$ thin films holds considerable promise for stabilizing targeted electronic phase diagrams and embedding them in technological applications. However, to date, the electronic structure of WO$_3$ thin films is experimentally unexplored, and only characterized by numerical calculations. Underpinning such properties experimentally would be important to understand not only the collective behavior of electrons in this transition-metal oxide, but also to explain and engineer both the observed optical responses to carriers' concentration and its prized catalytic activity. Here, by means of tensile strain, we stabilize WO$_3$ thin films into a stable phase, which we call pseudo-tetragonal, and we unveil its electronic structure by combining photoelectron spectroscopy and density functional theory calculations. This study constitutes the experimental demonstration of the electronic structure of WO$_3$ thin-films and allows us to pin down the first experimental benchmarks of the fermiology of this system.

---

Controlling the electronic properties of quantum systems allows us to realize technological applications with improved performance, stability, and durability, as well as significantly lower dissipation [1–3]. This is particularly relevant for *5d*-based transition metal oxides, which might provide a platform for integration into existing technology, with improved current densities, enhanced electrochromic and photovoltaic responses, and reduced switching energies [4–12]. Therefore, understanding the electronic structure of quantum systems is a crucial task, especially for newly synthesized materials, and it allows to pin down the hallmarks that describe their conductivity, their Fermi surfaces, and the relationship of the latter with symmetries and crystal structure.

Here, by using pulsed laser deposition (PLD) [13–15], we exploit epitaxial strain to synthesize a thermally stable phase in thin films of the *5d* compound WO$_3$ (on a LaAlO$_3$ substrate, LAO) and, by using angle-resolved photoelectron spectroscopy (ARPES), we unveil the electronic structure and properties, which describe the Fermi surface. The experimental demonstration of the electronic properties of $WO_3$ is a real milestone in the field of transition metal oxide, and its synthesis in thin films fashion into a stable phase is of great importance for gas sensor applications [16], water splitting [17], memory devices [18], high temperature diodes [19, 20], photodetectors [21, 22], for faster and more efficient electronics [4–12], and for low-dissipation Rashba ferro- and anti-ferroelectrics, for which $WO_3$ has been theoretically proposed as a candidate system [24]. Importantly, all these properties, which are relevant for both solid state optical devices and catalysis, are intimately connected to the electronic structure of this compound, the knowledge of which is so-far experimentally unexplored. We uncover the reference experimental benchmarks for the electronic band structure of $WO_3$ thin films, that despite the numerous studies which rely on it [23–31], is still lacking. In addition, by combining the experimental results with theoretical calculations, we report the existence of large distortion-induced band splitting, further enhanced by the spin-orbit coupling (SOC), shedding light on the mechanisms by which orbital hybridization occurs.

$WO_3$ thin films were grown by PLD at the NFFA facility [15]. The growth was performed at ~1000 K in an oxygen background pressure of $10^{-3}$ mbar (the typical deposition rate was 0.07 nm per laser shot). All of the investigated samples were grown on (001)-oriented LAO substrates. The ARPES measurements were performed *in-situ* by using a Scienta DA30 hemispherical analyzer with energy and momentum resolutions better than 15 meV and 0.02 Å$^{-1}$, respectively. The density functional theory (DFT) calculations were carried out within the CRYSTAL17 code [32] based on a linear combination of localized basis functions and the B1-WC hybrid functional [33]. To estimate/quantify the spin-orbit coupling (SOC), we also used the ABINIT code [58, 59], as described in the methods section.

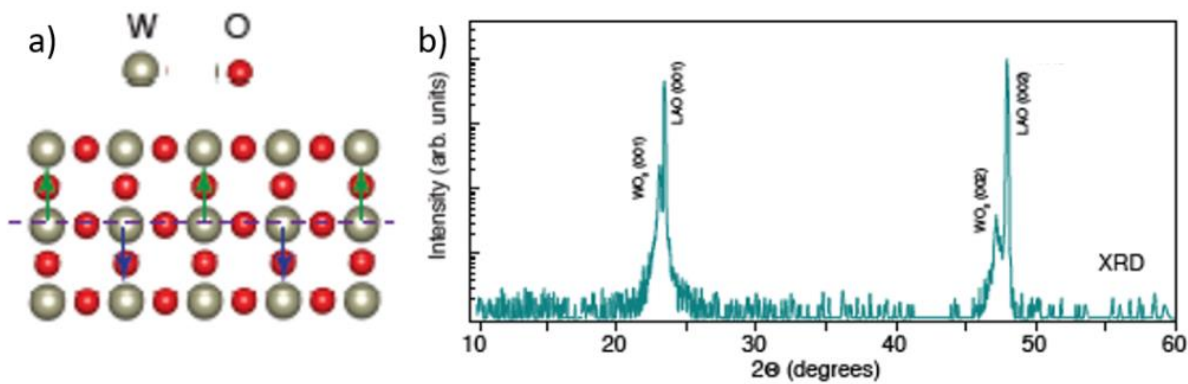

FIG. 1. (a) $WO_3$ tetragonal structure showing the out-of-plane opposite displacement of W and O atoms against one another. This phenomenology is also known as out-of-plane antipolar motion. (b) XRD θ–2θ scan of a $WO_3$ film grown on LAO (film peaks are indeed by considering a tetragonal structure).

WO$_3$ can be seen as an ABO$_3$ cubic perovskite with missing cation. It has however never been observed in the reference cubic structure, which exhibits various unstable phonon modes including antipolar motion of W against O in various directions (X$_5^-$ and M$_3^-$, see Fig. 1a) and oxygen octahedra rotations with different tilt patterns (M$_3^+$ and R$_4^+$) [28, 40]. Accordingly, in the bulk form, WO$_3$ undergoes several phase transitions as function of temperature: between 1300-1500 K, its structure is tetragonal (space groups P4/nmm, and P4/ncc) [34, 35], at 1000 K it becomes orthorhombic (Pbcn) [34–36], at room temperature monoclinic (P2$_1$/n) [35, 36], at 273 K triclinic (P-1) [37], and finally, at 200 K, it goes into a second monoclinic phase (P2$_1$/c) [28, 38-40], with no further transitions down to 5 K. This implies that this monoclinic phase is the ground state of bulk WO$_3$ [28].

The lattice parameters of the room-temperature monoclinic P2$_1$/n phase of bulk WO$_3$ are a = 0.732 nm, b = 0.756 nm, and c = 0.772 nm [35,36] and these remain very similar in the Pbcn, P-1 and P2$_1$/c phases. Taking into account the cell-doubling in all three directions, these lattice constants correspond to lattice spacing of about 0.366 nm along a, 0.378 nm along b, and 0.386 nm along c. In the P4/nmm phase, the lattice spacing is instead of about 0.375 nm along a and b and 0.392 nm along c. With respect to the LAO substrate - characterized by an in-plane pseudo-cubic lattice parameter of 0.379 nm - an epitaxial tensile strain is therefore expected for all phases. In our work, the stabilization of a structural phase with tetragonal metric at room temperature has been confirmed by the X-ray diffraction (XRD) data of Fig. 1b: from the (002) Bragg reflection, a *c*-axis parameter of 0.385 nm has been measured. The c value of this pseudo-tetragonal phase apparently matches that of the bulk P2$_1$/n and other low-temperature bulk phases. This is however surprising in view of the tensile epitaxial-strain conditions, expected to produce a significant contraction along c and better suggests that our film could adopt a P4/nmm type of structure. It has nevertheless been shown that changing the oxygen pressure during PLD growth has a major impact on the film out-of-plane lattice constant [41,30].

According to the report by Ning *et al* [41], decreasing the oxygen pressure (from 0.2 mbar to 6×10$^{-3}$ mbar) during WO$_3$ growth on LAO causes oxygen vacancies to emerge in the film, which results in an increase in the c parameter by up to 5%. As in other oxides [42,43], oxygen vacancies appear to be preferentially located at specific position of the perovskite structure rather being randomly distributed within the materials [42-44]. In this case, oxygen vacancies at apical positions of the WO$_3$ octrahedra result into an enlargement of the out-of-plane lattice parameter while a and b lattice parameters are unaffected [41]. This out-of-plane lattice expansion due to the presence of oxygen vacancies is often referred to as chemical strain [45,46] and could compensate the contraction produced by the tensile conditions imposed by the substrate. The measured value of c = 0.385 nm obtained from our experiment at 10$^{-3}$ mbar is in fact in very good agreement with the trend of c lattice parameter variation vs oxygen pressure reported in Ref. [41] for the P2$_1$/n phase, suggesting that our pseudo-tetragonal film might in fact better adopt either that structure or that of one of the similar low-temperature phases.

In order to clarify this issue, we have adopted an atomistic approach and performed DFT calculations. To determine the theoretical ground state of the WO$_3$ film, we focused on the six phases which are observed experimentally in the sequence of structural phase transitions of bulk

WO$_3$ [34-40], and explored their energy gain under tensile strain. Starting from the atomic positions of their fully relaxed bulk structures, we fixed their a and b lattice parameters to the pseudo-cubic a$_{LAO}$ = 0.379 nm, while relaxing the c parameter. Our calculations suggest that the theoretical ground state of the film should be the strained monoclinic P2$_1$/n phase with c = 0.738 nm. This result is in line with previous studies of stoichiometric WO$_3$ films, for which the c parameter was measured to be 0.733 nm [41, 47-48] and the structure of the film identified as similar to the monoclinic P2$_1$/n phase [41]. This result is however questioned by the observed c = 0.77 nm in our XRD.

As previously discussed, our films grown at low oxygen pressure are oxygen deficient. This was further confirmed by our photoemission data, which report metallic character for the samples, with the Fermi level crossing the conduction band, instead of an insulating behavior expected for the stoichiometric phase of WO$_3$. The oxygen vacancies then give rise to a chemical strain, enlarging artificially the c parameter. Following what was done in Ref. [46], the sub-stoichiometric character of our WO$_3$ film was then simulated by treating oxygen vacancies as a strain constraint in the out-of-plane direction. Accordingly, in addition to a = b = a$_{LAO}$, the c parameter was fixed to the experimental value (c = 0.77 nm). The energy gain diagram presented in Fig. 2a indicates that, in this specific case, the most stable configuration corresponds to the Pbcn structure. This suggests that our pseudo-tetragonal films might likely adopt that structure, which will be further confirmed later from the inspection of the electronic properties.

By using AMPLIMODE software [49], we performed symmetry-adapted mode analysis to identify the distortions, which play a major role in the stabilization of such a Pbcn strained phase. It can be characterized (see Fig. 2a) by (i) octahedra rotations (R$_4^+$ and M$_3^+$ modes) with tilt pattern a$^0$b$^+$c$^-$ in Glazer's notation [50], (ii) an antipolar motion along y (X$_5^-$ mode), (iii) a small contribution of a bending mode (X$_5^+$), and, finally (iv) an antipolar motion along the z and x axes (M$_3^-$ mode), where the x component of the M$_3^-$ mode appears through anharmonic coupling [40]. This is in contrast with the P2$_1$/c ground state of bulk WO$_3$ that arises from the contributions of (i) R$_4^+$ with tilt pattern a$^-$a$^-$c$^-$, (ii) antipolar motion along the z axis (M$_3^-$), and (iii) antipolar motion with the same amplitude along x and y axes (X$_5^-$).

Remarkably, we note that the pseudo-tetragonal thin films are incredibly resilient and their structure survives within a large temperature range, i.e. from room temperature (as demonstrated by XRD) down to, at least, 77 K (as confirmed by ARPES). This indicates that WO$_3$ on LAO has high structural and thermal stability and that the substrate can freeze the overgrown thin layers and make them robust against temperature variations. This is in contrast to the bulk behavior, in which orthorhombic (or tetragonal) phases have never been found at low temperatures but only at temperatures as high as 800 K [34-36, 52-53]. Again, this result points at the importance of epitaxial strain in realizing films with enhanced thermal stability compared to the bulk counterpart.

In order to understand the role of the crystal structure in the electronic properties of this compound, we performed ARPES with in-vacuum transfer, without exposing the samples to air. First of all, we notice that the tetragonal metric of the WO$_3$ films is also reflected in the symmetries of the reciprocal space, namely in the symmetry of the Fermi surface (See Fig.3a). The latter shows a Fermi level crossing along $k_x$ and $k_y$ of 0.45 Å$^{-1}$, as indicated in Fig. 3a. From the Γ to the X points of the Brillouin zone (Fig. 3a), we did not observe any appreciable change in the Fermi surface volume within our experimental resolutions, however, an overall different shape is visible, as expected for this system, which electronically speaking still behaves bulk-like for the ARPES probing depth. To locate the high symmetry positions along the $k_z$ direction, we performed photon-energy dependent scans (See Supplementary information for $k_z$ vs photon energy dependence) and we show them in Fig. 3b. Here, we see 'hot spots' of spectral intensity at several $k_z$ values. This repeating behavior helps us to fix the $k_z$ corresponding to the high symmetry points of the Brillouin zone, namely X and Γ, and allows us to make an estimate of the c-axis from ARPES. With an inner potential of 11 ± 3 eV, we obtain $c_{ARPES}$ = 0.77 nm, in agreement with the results from XRD. We notice that ARPES gives exactly twice the XRD value, indicating that the unit cell has a doubling, here revealed by the resonant behavior of the spectra. The structural distortions of the pseudo-tetragonal phase lead to large energy splittings, in contrast to the Pbcn , P2$_1$/n and P2$_1$/c bulk phases (see supplementary information Fig. 2). In high-temperature tetragonal bulk WO$_3$, the $d_{xz}$ and the $d_{yz}$ orbitals are degenerate at the center of the BZ (see supplementary information Fig. 2a). However, in the pseudo-tetragonal thin film, the orbital degeneracy is removed, resulting in a energy splitting that can can be resolved by ARPES measurements and takes the experimental value of 100 meV (Fig. 2 d,e). This splitting is consistent with our calculations in the strained Pbcn phase, although the computed value takes a smaller value of as 60 meV (see Fig. 2 b).

By including SOC, which is expected to be relevant for *5d* orbitals, this discrepancy finds a solution; our calculation in the strained Pbcn phase reproduce an energy splitting of ≈ 90 meV between $d_{yz}$ and $d_{xz}$, in perfect agreement with ARPES (Fig. 2 c). This emphasizes the SOC's critical role in hybridizing the orbitals in 5d oxides. However, our results suggest that the effect of structural distortions is greater than the SOC. In the WO$_3$ film, the amplitude of the antipolar distortion along y is greater than along x, as a result the upshift of the $d_{yz}$ energy level is larger than $d_{xz}$. This is because an increase in the y-direction antipolar distortion results in a stronger overlap between the O 2p$_y$ and W 5d$_{yz}$ orbitals [51], which in turn causes an upshift in the related antibonding energy level. Thus, the splitting between $d_{yz}$ and $d_{xz}$ in this pseudo-tetragonal phase (shown in Fig. 2b,c) is caused by a proper balance between the amplitude of the antipolar motions along y (X$_5^-$ mode), and x (M$_3^-$ mode). Note that this splitting, with the $d_{yz}$ located at a higher energy level than the $d_{xz}$, is absent in all the bulk phases, including the Pbcn and P2$_1$/n phases where the x component of the M$_3^-$ mode is negligible, or, in the bulk form of the ground state where the amplitude of the antipolar motion along y and x are almost the same (see supplementary information Fig.2). More importantly, as shown in Fig. 3 of the supplementary information, this is not the case in any of the similar low-temperature phases under strain, providing additional evidence that our pseudo-tetragonal film adopts a strained Pbcn structure.

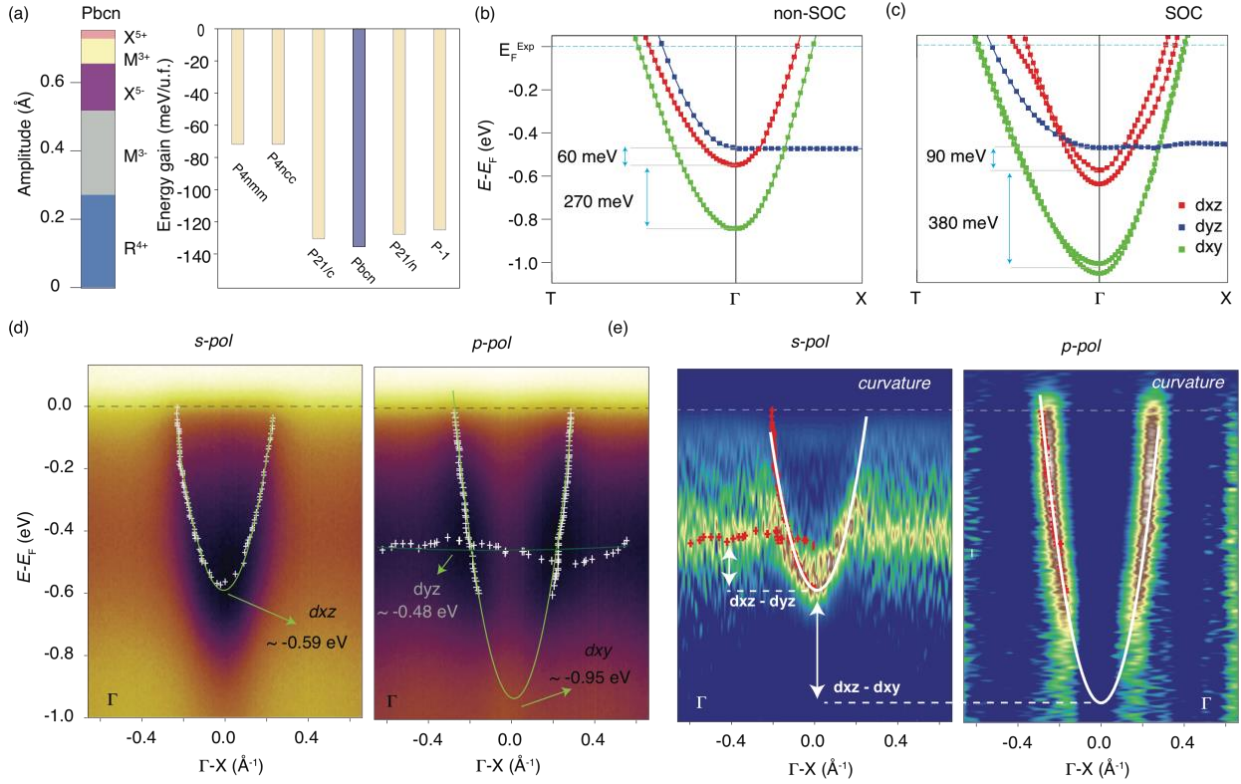

FIG. 2. (a) The symmetry-adapted mode decomposition pseudo-tetragonal thin film with Pbcn strained structure of $WO_3$ (left) and the DFT calculated energy gain of the six phases which are observed experimentally in the sequence of structural phase transitions of bulk $WO_3$ with respect to the cubic phase (right), obtained by fixing the lattice parameters to the experimental ones to take into account the tensile strain as well as the strain constraint in the out-of-plane direction to account for the sub-stoichiometric nature of our $WO_3$ film. The DFT electronic structure of pseudo-tetragonal thin film with Pbcn strained structure is shown in (b) and (c), without and with SOC, respectively. The relevant energy splitting is captured by DFT and the orbital mixing induced by the SOC enhances further the energy separation. The Fermi level in the DFT calculations has been aligned to the experimental value by rigidly shifting the calculated bands. As shown in supplementary Fig.3, where the electronic structures of the similar low-temperature strained phases of $WO_3$ are presented, our ARPES data are only compatible with the strained Pbcn structure. Only in this structure are the orbitals in the same sequence, and $d_{xz}$ and $d_{xy}$ are split considerably. (d) ARPES data along the Γ-X direction are shown, showing a good agreement with the calculations. The minima of the $d_{yz}$ and $d_{xy}$ bands are shown, as well as an average value for the energy at which the dispersion-less $d_{xz}$ is located. (e) The ARPES curvature plots are shown to better visualize the energy states and their relative separation, indicated by the white arrows for the bands relative to each other's.

A second splitting is also observable in the DFT results between the $d_{xz}$ and $d_{xy}$ orbitals and it is estimated to be ≈ 380 meV after inclusion of SOC (see Fig. 2c). Our calculations indicate that, octahedra tilting ($M_3^+$ and $R_4^+$ modes) with W-O-W angle deviations from 180°, also involves tuning the overlap of orbitals in this case [54]. From ARPES (Fig. 2d and Fig. 2e), it is more challenging to make a straightforward comparison with the DFT results, because the $d_{xy}$ band has strong matrix elements which suppress its intensity near the centre of the BZ [55, 56]. However, we can extrapolate the minimum by fitting the data, obtaining a $d_{xz}$-$d_{xy}$ separation of ≈ 400 meV, which is also in close agreement with the calculated value. Thus, the structural distortions are very important in defining the electronic properties of $WO_3$ and the strain is crucial for stabilizing the pseudo-tetragonal phase observed here.

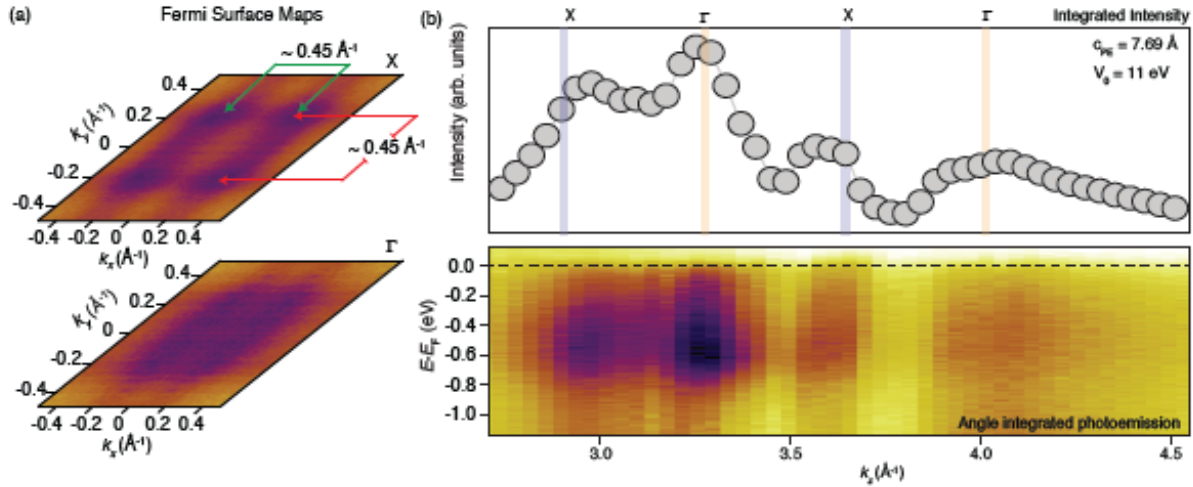

FIG. 3. (a) Fermi surfaces collected for the first X and Γ points in (b). The pattern observed in the spectra is consistent with a tetragonal unit cell structure. For reference, we have also shown the calculated iso-energetic cuts across the WO3 conduction band, which show a good agreement with the experiment (Supplementary Fig.4.) (b) Resonant photoemission signal. The inner potential was also obtained by fitting the data and found to give the best agreement at V0 = 11 eV. The procedure is explained in the supplementary information and within our resolutions V0 between 8 eV and 14 eV did not show any substantial change in the c-axis value (See also supplementary Fig.1).

In conclusion, we report the existence of a new phase in $WO_3$, which we call pseudo-tetragonal but reveals in fact a strained Pbcn phase. This phase observed in films grown at low oxygen pressures differs from the $P2_1/n$ phase previously reported in stoichiometric films. It accommodates antipolar distortions along all three axes. Such distortions are important to understand the vibrational modes and the electronic properties of this system. By combining XRD, TEM, DFT calculations, and ARPES we determine the role and consequences of the structural distortions on the $WO_3$ electronic structure, experimentally revealing a band splitting as large as 400 meV between the $d_{xz}$ and $d_{xy}$ orbitals, and 100 meV between the $d_{yz}$ and $d_{xz}$ orbitals, reminiscent of the proper balance between the amplitude of the $M_3^-$ and $X_5^-$ antipolar modes in different directions [28, 40, 55–57]. Finally, we show a large thermal stability for the grown films, and we demonstrate that the SOC plays a sizeable role for the interpretation of the electronic behavior of $WO_3$. Our work not only motivates the use of strain to realize novel structural phases in binary *5d* oxides, but also shows how to use it to tune their orbital degrees of freedom.

**Acknowledgements**
Ph.G. acknowledges support from F.R.S.-FNRS through the PDR project PROMOSPAN. The authors greatly acknowledge Prof. M. Gennou for the useful discussions and insights. Calculations were performed using CECI supercomputer facilities funded by the FRS-FNRS (Grant No. 2.5020.1), the Tier-1 supercomputer of the Federation Wallonie- Bruxelles funded by the Walloon

Region (Grant No. 1117545), and the computing facilities of the Flemish Supercomputer Center. F.M. acknowledges Prof. Pascal Turban and Francine Solal for the useful insights and helpful discussions. F.M. acknowledges also the SoE action of pnrr, number SOE_0000068.

## Methods

### DFT details

To approximate the Brillouin zone, integration over 8x8x8 k-point meshes for the cubic symmetry, or meshes with equivalent sampling for other phases (e.g., meshes of 6x6x8, 6x6x4, and 4x4x4 for the P4/nmm, P4/ncc, and Pbcn phases, respectively) were used. In the CRYSTAL17 code [32], the Self-Consistent-Field (SCF) convergence's tolerance on change in total energy was set to $10^{-10}$ Hartrees. The geometry optimization was performed by employing a quasi-Newton approach with a BFGS Hessian scheme so that a specific space group symmetry was preserved for each structure during the structural relaxations. The root mean square of the gradient and displacements were converged to less than $5\times10^{-5}$ Hartrees/Bohr and $10^{-3}$ Bohrs, respectively. We also used the ABINIT code [58, 59] with a plane-wave basis set and the LDA functional with Perdew-Wang's parametrization [60], in order to include spin-orbit coupling (SOC) for the electronic band structures. In this case, the electronic wave functions were expanded in plane waves up to an energy cutoff of 60 Hartrees and the electronic self-consistent calculations were converged until the difference of the total energy is smaller than $10^{-9}$ Hartrees.

# Supplementary Information: Unveiling the electronic structure of pseudo-tetragonal WO$_3$ thin films

F. Mazzola,[1,2,3,*] H. Hassani,[4,5,*] D. Amoroso,[4] S.K. Chaluvadi,[1] J. Fujii,[1] V. Polewczyk,[1] P. Rajak,[1] Max Koegler,[1] R. Ciancio,[1] B. Partoens,[5] G. Rossi,[6,1] I. Vobornik,[1] P. Ghosez,[4] and P. Orgiani[1]

[1]*CNR-IOM TASC Laboratory, Area Science Park, I-34149 Trieste, Italy*
[2]*Department of Molecular Sciences and Nanosystems,*
*Ca Foscari University of Venice, 30172 Venice, Italy*
[3]*Istituto Officina dei Materiali (IOM)-CNR, Laboratorio*
*TASC,Area Science Park, S.S.14, Km 163.5, 34149 Trieste,*
*Italy*
[4]*Theoretical Materials Physics, Q-MAT, CESAM, Université de Liège, B-4000 Lige,*
*Belgium* [5]*Department of Physics, University of Antwerp, Groenenborgerlaan 171, 2020 Antwerp,*
*Belgium*[6]*University of Milano, I-20133 Milano, Italy*
(Dated: April 11, 2023)

[*] These authors contributed equally to this work.

**ARPES $k_z$ plot procedure and measurements**

In order to access the electronic dispersion along the $k_z$ direction, as shown in Fig. 3b of the main text, we collected ARPES energy-k dispersions at several photon energies. Being the photon energy and the $k_z$ directly proportional, we can readily obtain the dispersion along the out-of-plane direction of the Brillouin zone. In particular:

$$(1) \quad k_z = \frac{1}{\hbar}\sqrt{2m(E_{kin}cos^2\theta + V_0)}$$

where, m is the electron mass, $E_{Kin}$ the electrons' kinetic energy, $\theta$ is the polar angle, and $V_0$ is the inner potential. Thus, when varying photon energy, one is able to detect the electronic structure of solid state systems along the $k_z$ direction by verifying how the bands change. As an example, we show the linear vertical polarization data for $WO_3$ collected for some photon energies in Fig.1.

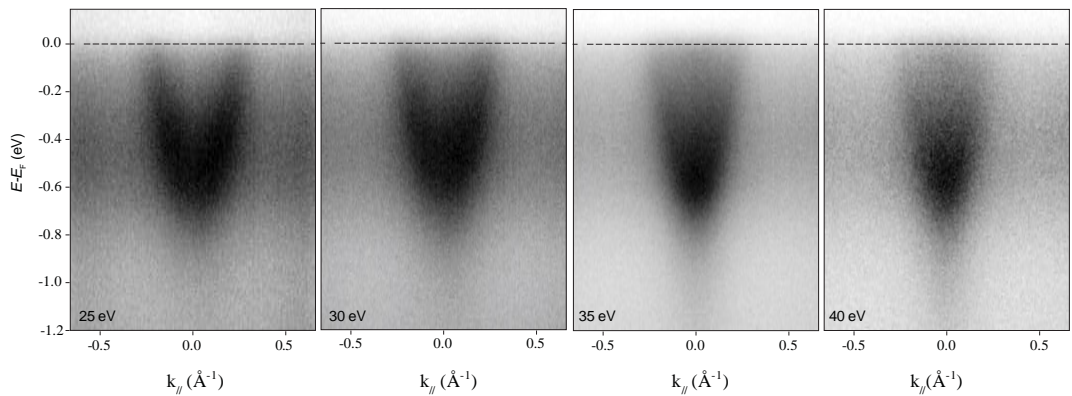

FIG. 1. Electronic structure of $WO_3$ collected by ARPES at various photon energies, showing changes which can be attributed to a combination of matrix elements and to the three-dimensional character of the material. These plots relate directly to figure 3b of the main text.

We notice that the ARPES measurements were possible because the samples had states at the Fermi level. This is a known situation which occurs for many oxides, in which even a mild chance in oxygen balance could affect the electronic properties, without changing necessarily the structure [1, 2].

**ARPES fitting details**

The ARPES spectra have been fitted by using Lorentzian curves convoluted by a Gaussian to account for the resolutions of the instrument, which were better than 15 meV and 0.02 $^{-1}$, for energy and momentum respectively. The data were fitted by using energy-distribution and momentum-distribution curves to properly account for the curvature of the various features: for example, at the minimum of the parabolic dispersions, energy distribution curves capture better the energy dispersion. At the Fermi level, instead, momentum distribution curves are more suited to give an accurate determination of the k-position of the electronic states.

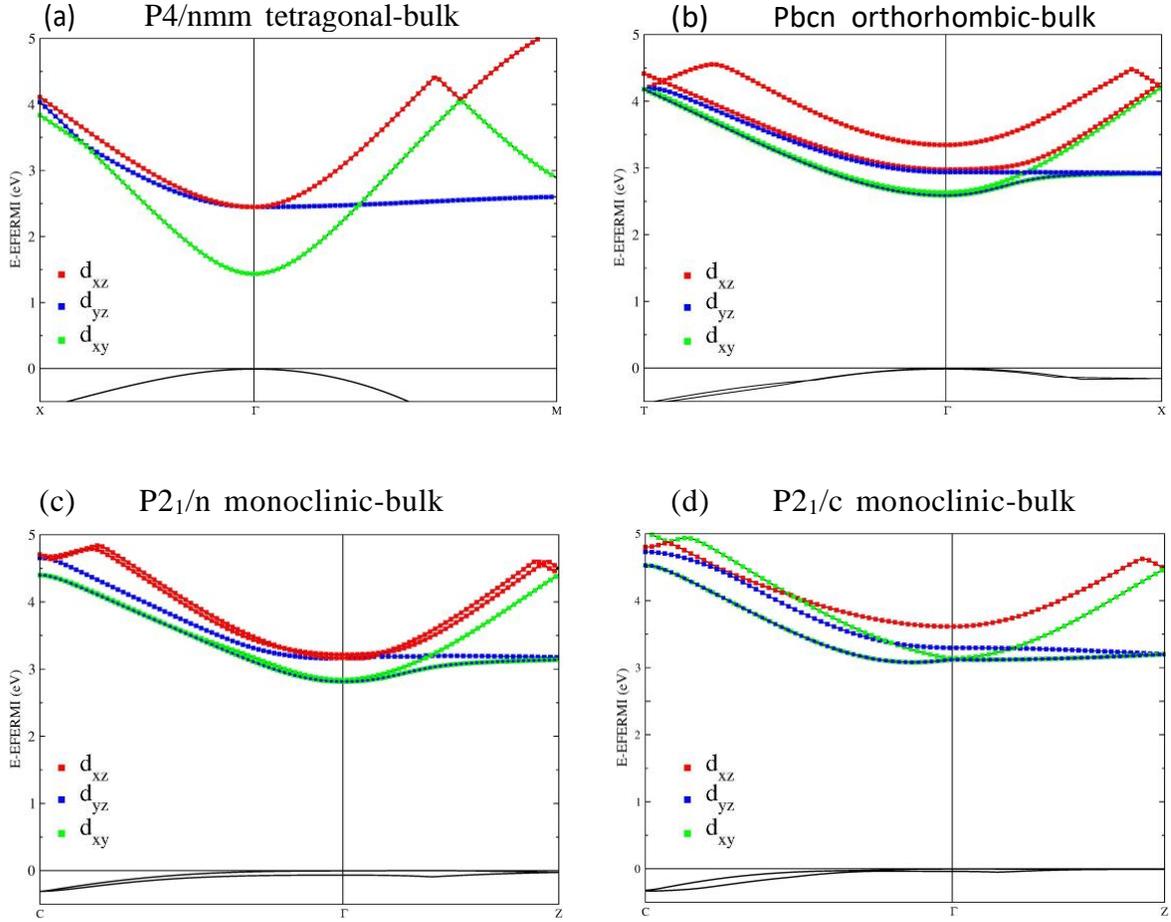

FIG. 2. (a) DFT electronic structure of the P4/nmm tetragonal phase in the bulk form, (b) the orthorhombic Pbcn phase, (c) the room-temperature P2$_1$/n monoclinic phase, and (d) the ground state P2$_1$/c monoclinic phase. The large splittings in the WO$_3$ thin films, with the same electronic level arrangement, are absent in all the phases of bulk form.

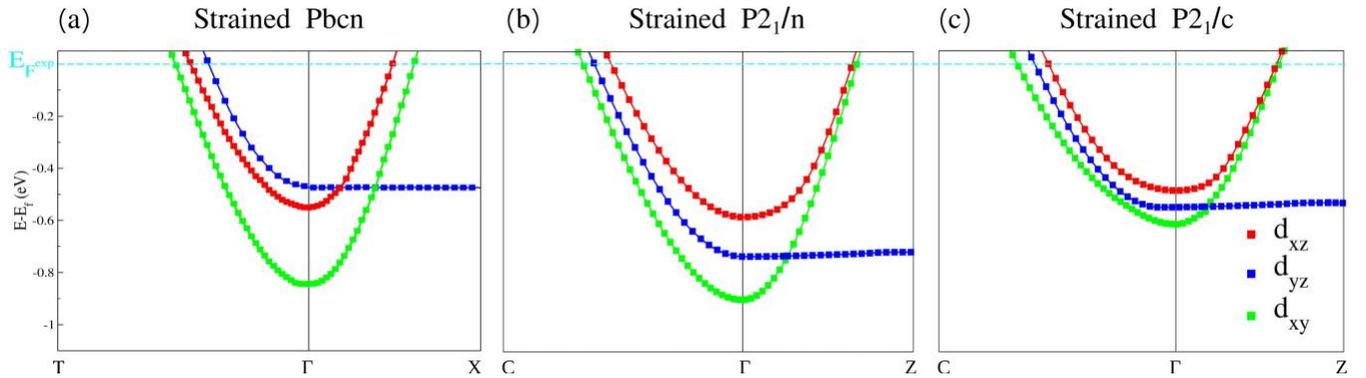

FIG. 3. DFT electronic structure (without SOC) of the pseudo-tetragonal thin film with (a) strained Pbcn structure, (b) strained P2$_1$/n structure, and (c) strained P2$_1$/c structure. The Fermi level in the DFT calculations has been aligned to the experimental value by rigidly shifting the calculated bands. It can be seen that our ARPES data are only compatible with strained Pbcn structure. Only in this structure are the orbitals in the same sequence, as well as the large splitting of d$_{xz}$ and d$_{xy}$, which is reminiscent of the correct balance of the amplitude of the $M_3^-$ and $X_5^-$ antipolar modes in various directions.

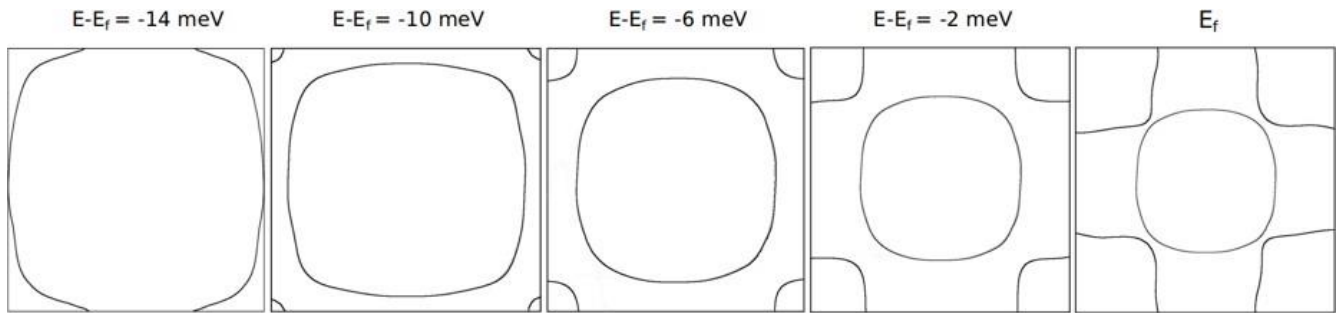

FIG. 4. DFT calculated iso-energetic cuts, showing a good agreement with the experimental Fermi surface in Fig. 3 (a) of the main text (The $E_f$ panel can be used for the comparison).